\documentclass[12pt]{iopart}
\usepackage[T1]{fontenc} % may fix type 3 fonts problem
\usepackage{ae,aecompl}

\usepackage{amssymb}
\usepackage{graphicx}
\usepackage{dcolumn}
\usepackage{bm}
\usepackage{booktabs}
\usepackage{ctable}

\begin{document}
\title[Storage-based modelling of heterocyst patterning in cyanobacteria]{A storage-based model of heterocyst commitment and patterning in cyanobacteria} 

\author{Aidan I Brown and Andrew D Rutenberg}
\address{Department of Physics and Atmospheric Science, Dalhousie University, Halifax, Nova Scotia, Canada B3H 1Z9}
\ead{\mailto{aidan@dal.ca}, \mailto{andrew.rutenberg@dal.ca}}

\date{\today}

%%PB max 200 words
\begin{abstract}
When deprived of fixed nitrogen (fN), certain filamentous cyanobacteria differentiate nitrogen-fixing heterocysts.  There is a large and dynamic fraction of stored fN in cyanobacterial cells, but its role in directing heterocyst commitment has not been identified. We present an integrated computational model of fN transport, cellular growth, and heterocyst commitment for filamentous cyanobacteria. By including fN storage proportional to cell length, but without any explicit cell-cycle effect, we are able to recover a broad and late range of heterocyst commitment times and we observe a strong \emph{indirect} cell-cycle effect. We propose that fN storage is an important component of heterocyst commitment and patterning in filamentous cyanobacteria. The model allows us to explore both initial and steady-state heterocyst patterns. The developmental model is hierarchical after initial commitment: our only source of stochasticity is observed growth rate variability.  Explicit lateral inhibition allows us to examine $\Delta$\emph{patS}, $\Delta$\emph{hetN}, and $\Delta$\emph{patN} phenotypes. We find that $\Delta$\emph{patS} leads to adjacent heterocysts of the same generation, while $\Delta$\emph{hetN} leads to adjacent heterocysts only of different generations. With a shortened inhibition range, heterocyst spacing distributions are similar to those in experimental $\Delta$\emph{patN} systems. Step-down to non-zero external fixed nitrogen concentrations is also investigated.
\end{abstract}
\pacs{87.17.Aa, 87.17.Ee, 87.18.Fx, 87.18.Tt}
%87.17.Aa Cell processes modeling, computer simulation
%87.17.Ee Cell processes growth and division
%87.18.Fx Biological complexity multicellular phenomena
%87.18.Tt Biological complexity noise in biological systems
\noindent{\it Keywords\/}: cyanobacteria, heterocyst commitment, developmental pattern, computational model,  nitrogen storage, lateral inhibition

\maketitle
%%%%%%%%%%%
%%%%%%%%%%%
\section{Introduction}

Certain filamentous cyanobacteria are model multicellular organisms that, under conditions of low exogenous fixed nitrogen (fN), terminally differentiate regularly spaced heterocysts that fix nitrogen for interspersed clusters of photosynthetic vegetative cells. While much is known about the genetic regulation of heterocyst development \cite{flores10, kumar10}, the mechanism by which certain cells are selected to become heterocysts has nevertheless remained mysterious. 

There are prompt signs of nitrogen deprivation after the step-down of fN, such as the activation of nitrate assimilation genes \cite{wolk96} and the increase of 2-oxoglutarate levels \cite{laurent05} within 1h. After this, there is a long delay until 8h to 14h after the fN step-down, when the first generation of cells commit to become heterocysts \cite{yoon01}. Rapid diffusion of calcein, a non-native fluorescent reporter, has been observed in filamentous cyanobacteria \cite{mullineaux08}. This suggests that the diffusion of small molecules carrying fN along the filament will also be rapid \cite{brown12}, and that commitment may be influenced by more than the depletion of diffusible fN alone. 

Suggestions that heterocyst differentiation is strictly dependent on the cell cycle \cite{mitchison76, sakr06} have not been borne out by experiment \cite{asai09, toyoshima10}. However, unicellular studies show that cell size can affect cell fate decisions without explicit cell-cycle effects, for example in yeast meiosis \cite{nachman07} or in phage lambda infection \cite{stpierre08}. While postulated mechanisms of cell size effects vary, they include nutrient availability \cite{stpierre08}. In the multicellular cyanobacterial system, substantial stores of fN are available.  Cyanophycin and phycobiliprotein represent up to 20$\%$ \cite{li01} of protein and 60$\%$ \cite{bogorad75} of soluble protein, respectively, and are depleted under fN limitation \cite{allen84}. Could there be a role for fN storage, involving cell size, in heterocyst commitment?   

Buikema and Haselkorn \cite{buikema93} suggested that uneven accumulation of cyanophycin would cause some cells to starve first, leading to commitment variability, but reported no evidence of uneven accumulation of cyanophycin along the filament. However, a uniform concentration (per unit volume) of fN storage combined with the natural diversity of cell sizes in growing filaments would lead to uneven total fN storage per cell along the filament --- because length and hence volume would vary from cell to cell.

We explore the hypothesis that an initially uniform concentration of fN storage could explain the timing and variability of initial heterocyst commitment in \emph{Anabaena} PCC 7120. We do this within the context of a computational model of filament growth and development that allows for fN storage, transport, and consumption by cell growth.  We impose a deterministic, or hierarchic \cite{buganim12}, developmental schedule that committed cells follow to become heterocysts. This deterministic schedule ensures that the only stochastic event in heterocyst development is the time of commitment. In particular the model allows for no intrinsic variability in lateral inhibition due to {\em patS} \cite{yoon01, yoon98} and {\em hetN} \cite{callahan01}. Developmental and systems-biology studies of cyanobacterial filaments need to account for the population context of cells \cite{snijder09}, and the model indicates that the total fN storage of each cell may be an important source of variability determining which cells differentiate into heterocysts.

We can use the model to examine how the heterocyst pattern changes with time --- from the initial commitment to the spacing distribution at 24h and 48h after fN step-down.  We find that uniform storage together with a natural distribution of cell lengths is enough to provide a strong {\em indirect} cell-cycle effect on committed heterocysts, with shorter cells favoured for heterocyst differentiation \cite{mitchison76, sakr06}.  We obtain the observed 8-14h timing and range of commitment seen in experiments \cite{yoon01}, and predict that both storage fraction and growth rate should control overall commitment timing.

The model also recovers the multiple-contiguous heterocyst (Mch) phenotype and timing of both  $\Delta${\em patS} \cite{yoon01, yoon98} and $\Delta${\em hetN} \cite{callahan01} knockouts --- and predicts that clusters of heterocysts seen in $\Delta${\em hetN} filaments will always include distinct generations of heterocysts, and so will be qualitatively distinct from heterocyst development in $\Delta${\em patS} filaments. The model is used to show that the heterocyst patterning effects similar to those of $\Delta$\emph{patN} \cite{risser12} can be obtained by simply reducing the inhibition range of {\em patS} and {\em hetN}.

Some heterocystous cyanobacterial filaments fix nitrogen in vegetative cells under anoxic conditions, and yet still differentiate heterocysts in a pattern \cite{thiel01}. Under the assumption that such vegetative fixation provides a background level of fN that is not quite sufficient for maximal vegetative growth, we also investigate heterocyst differentiation when the nitrogen steps down to nonzero external fixed nitrogen concentrations.  These conditions could be accessed in developmental studies using flow chambers. 

%%%%%%%%
%%%%%%%%
\section{Computational Model}
\label{sec:model}

An early quantitative fN transport and consumption model by Wolk \emph{et al} \cite{wolk74} considered how fN would spread from a heterocyst to surrounding vegetative cells, but did not include distinct cells or stochasticity and assumed that fN was always rate-limiting for growth. Allard \emph{et al} \cite{allard07} considered dynamic heterocyst placement in growing, stochastic filaments of cells with fN transport but only used periplasmic transport without inhibitors. Their fN dynamics was adapted for cell-cell transport by Brown and Rutenberg \cite{brown12}, who demonstrated that a constant cell growth rate in the presence of fixed nitrogen was consistent with observed nitrogen distributions between two heterocysts.  This model was extended to the filament level \cite{brown12b} to recover heterocyst spacing distributions during steady-state growth. These models suggest that fN dynamics are important for steady-state heterocyst patterns, but could not address the initial heterocyst patterns observed soon after fN step-down since they did not explicitly include lateral inhibition due to, e.g., \emph{patS} and \emph{hetN}. 

Earlier modelling efforts showed that initial heterocyst patterns were not random \cite{meeks02} but were consistent with diffusible inhibitors \cite{wolk75}. These models did not include growth, fN transport, or distinct effects of \emph{patS} and \emph{hetN}. Recently, an integrated local model with static heterocysts has been presented \cite{zhu10} --- but without stochasticity it is unable to address heterocyst pattern formation at the filament level.

In this work we implement a stochastic computational model of the cyanobacterial filament, focusing on the model organism \emph{Anabaena} PCC 7120. The model has growth, division, and fN transport, along with dynamic heterocyst placement using storage-based commitment and explicit lateral inhibition of differentiation. Our quantitative heterocyst differentiation model is  similar in spirit to the qualitative proposal by Meeks and Elhai  \cite{meeks02} of a two-stage model of heterocyst differentiation with biased initiation based on the cell cycle followed by competitive resolution of differentiating clusters to isolated heterocysts. The crucial difference is that the model we propose includes local fN storage, which we use to initiate development instead of an explicit cell cycle effect.  Since we have growth, fN dynamics, and explicit lateral inhibition due to \emph{patS} and \emph{hetN} in a stochastic developmental model we can calculate both initial and steady-state heterocyst patterns and compare them with experimental distributions. 

%%%%%%%%%%%%%%%%%%%
\subsection{Fixed nitrogen transport}
The dynamics of freely diffusing fixed nitrogen, $N_i$, are given by
\begin{equation}
\label{eq:ndynamics}
	\frac{d N_i}{dt} =\Phi_i+D_I\rho_{efN}l_i-D_LN_i+G_i,
\end{equation}
where $\Phi_i\equiv\Phi_R(i-1)+\Phi_L(i+1)-\Phi_R(i)-\Phi_L(i)$ is the net diffusive flux into cell $i$. Each flux is the product of the local density $N_i/l_i$, where $l_i$ the length of cell $i$, and a transport coefficient $D_C$ which accounts for the flux magnitude and absorbs all geometric factors. We use $D_C=1.54$ $\mu$m s$^{-1}$ between two vegetative cells and $D_C=0.19$ $\mu$m s$^{-1}$ between a vegetative cell and a heterocyst \cite{brown12}.  The import from outside the filament is expected to be proportional to the number of transporters, which will be proportional to the cell surface area and thus the cell length for cylindrical cells. Import is then proportional to the cell length, the coefficient $D_I=2.9\times10^{-18}$m$^3$($\mu$m s)$^{-1}=1.7\times10^9$M$^{-1}$($\mu$m s)$^{-1}$, which has geometric factors included, and the external fN concentration $\rho_{efN}$ \cite{brown12b}. We also allow a small leakage, which is expected to be proportional to cytoplasmic density, $N/(Al)$, with $A$ the cross-sectional area, but also the number of transporters, which is expected to be proportional to cell length $l$. This results in a leakage term independent of cell length and proportional to the amount of fixed nitrogen $N$ and the coefficient $D_L=0.029/\pi$ s$^{-1}$ (geometric factors are absorbed) \cite{brown12b}. $G_i$ is a source/sink term and is described below.

%%%%%%%%%%%%%%%%%%%%%
\begin{table}
\centering
\caption{\label{tab:parameters}{\bf Model parameter definitions and values.} Shown are the standard values used. Further discussion is found in Sec.~\ref{sec:model}}
\begin{tabular}{lll}
\toprule
Variable&Description&Value/Eqn\\
\midrule
$N_i$&cytoplasmic fN amount of cell $i$&variable\\
$N_{Si}$&stored fN amount of cell $i$&variable\\
$\Phi_i$&diffusive flux from neighbouring cells to cell $i$&variable\\
$\Phi_{in}$&influx of fN from neighbours and outside filament&variable\\
$\rho_{efN}$&external fN concentration&varied\\
$D_C$&coefficient for cell to cell fN transport&1.54 $\mu$m s$^{-1}$\\
$D_I$&coefficient for fN import from external fN&2.9$\times10^{-18}$m$^3$($\mu$m s)$^{-1}$\\
$D_L$&coefficient for fN leakage&0.029/$\pi$ s$^{-1}$\\
$G_i$&source/sink term related to fN production/growth&variable\\
$G_{veg}$&cytoplasmic fN removed to support vegetative growth&Eqn.~\ref{eq:gveg}\\
$G_{het}$&heterocyst fN production&3.15$\times$10$^6$ s$^{-1}$\\
$l_i$&length of cell $i$&[$l_{min}$, $l_{max}$]\\
$l_{min}$&minimum length of cell&2.25$\mu$m\\
$l_{max}$&cell length triggering division&4.5$\mu$m\\
$T_i$&doubling time of cell $i$&[$T_{min}$, $T_{max}$]\\
$T_{min}$&minimum cell doubling time&$T_D-\Delta$\\
$T_{max}$&maximum cell doubling time&$T_D+\Delta$\\
$\Delta$&range of cell doubling times&4.5h\\
$T_D$&average cell doubling time&20h\\
$R_i$&growth rate of cell $i$&Eqn.~\ref{EQN:Ri}\\
$\hat{R}_i$&maximal growth rate of cell $i$&$l_{min}/T_i$\\
$g$&fN cost per unit length of growth&6.2$\times$10$^9$/$\mu$m\\
$f_g$&maximal growth rate fraction provided by stored fN&0.25\\
$f_s$&fraction of fN used to grow that is stored&0.3\\
$\tau_N$&delay after het commitment until HetN inhibition begins&varied\\
$\tau_S$&delay after end of PatS inhibition until fN production begins&varied\\
\bottomrule
\end{tabular}
\end{table}

%%%%%%%%%%%%%%%%%%%
\subsection{Cell growth and division}
Following \cite{brown12b}, we take \emph{Anabaena} PCC 7120 cells to have a minimum size of $l_{min}=2.25$ $\mu$m and a maximum size of $l_{max}=2l_{min}$ \cite{flores10, kumar10}. When a cell reaches $l_{max}$ it is divided into two cells of equal length, each of which is randomly assigned a new growth rate and half of the fN and fN storage of the parent cell. Daughter cells assigned different growth rates will have different lengths at later times and so lead to a natural population structure with diverse cell lengths. To select a new growth rate we first define a minimum doubling time $T_{min}=T_D-\Delta$ and a maximum doubling time of $T_{max}=T_D+\Delta$, with an average doubling time $T_D=20$h \cite{picossi05} and a range $\Delta=4.5$h \cite{brown12b} that preserves the reported experimental coefficient of variation \cite{allard07}.  A doubling time $T_i$ is randomly and uniformly selected from its range and converted to a maximal growth rate $\hat{R}_i=l_{min}/T_i$.

Single cell growth rates have not been reported in cyanobacteria. However the length increase of individual \emph{E. coli} cells is approximately linear in time \cite{reshes08}. Accordingly, our model assigns each individual cell a constant maximal growth rate. The actual growth rate, $R_i$, is at most $\hat{R}_i$ but is reduced if the local availability of fN is insufficient to accommodate maximal growth \cite{brown12b}:
\begin{equation}
	R_i =
  		\cases{
    			\hat{R}_i & $N_i > 0$\cr
    			min\left(\Phi_{in}/g+f_g\hat{R}_i,\hat{R}_i\right)& $N_i = 0$, $N_{Si}>0$\cr
    			min\left(\Phi_{in}/g,\hat{R}_i \right)  		& $N_i = 0$, $N_{Si}=0$\cr
   }
   \label{EQN:Ri}
\end{equation}
where $g$ is the fN cost per unit length of growth, $f_g$ is the fraction of the maximal growth rate $\hat{R}$ that can be accommodated by stored fN, and $N_S$ is the amount of stored fN. The first condition imposes that any cytoplasmic fN can accommodate the maximal growth rate; the second condition that local storage can accommodate up to $f_g \hat{R}_i$ of  growth; the last two conditions that any influx of fN from neighbouring cells and outside the filament, $\Phi_{in}$, will immediately accommodate some growth when $N_i=0$. For vegetative cells
\begin{equation}
G_i=G_{veg}=
		       \cases{
		                    -gR_i& $N_i>0$\cr
		                    -\Phi_{in}& $N_i=0$.\cr
		                   }
		                   \label{eq:gveg}
\end{equation}
For simplicity, the model allows for vegetative growth even for vanishingly small $N_i$.  A minimum threshold concentration for growth could be easily included, though we do not anticipate  significant changes as long as the threshold is small compared to the cellular fN growth requirement per unit length, $g$.  

%%%%%%%%%%%%%%%%%
\subsection{Fixed nitrogen storage}
Each cell $i$ has an amount of stored fN, $N_{Si}$, that provides fN for the cell to grow when the diffusible cytoplasmic fN, $N_i$, has run out. Storage is composed of cyanophycin and phycobiliprotein, and this storage is degraded upon nitrogen deprivation \cite{allen84}. Once heterocysts have developed, cyanophycin is replenished in both vegetative cells and heterocysts \cite{picossi04}.  In the model, a fraction $f_s$ of the fN incorporated by the cell will be storage (we use $f_s=0.3$, see Sec.~\ref{sec:fs}). This then gives us the fN storage dynamics of the $i$th cell:
\begin{equation}
	\frac{dN_{Si}}{dt} =
		  \cases{
	    f_s g R_i & $N_i>0$  \cr
 	   f_s\Phi_{in}-(gR_i-\Phi_{in})& $N_i = 0$, \cr
 	 }
	 \label{eq:storage}
\end{equation}
where $g$ is the fN cost per unit length of growth, $R_i$ is the growth rate of the $i$th cell, and $\Phi_{in}$ is the influx of fN into the $i$th cell from adjacent cells and from outside the filament. For simplicity, we assume that only enough stored fN for local growth needs is released in Eqn.~\ref{eq:storage}. This released fN is immediately taken up by local growth, and none is available for diffusion to other cells. Storage regenerates when diffusible fN is available, with all cells gaining the same amount of storage per unit length grown when $N_i>0$. A filament that has been grown indefinitely in excess fN will have cells with the same amount of storage per unit length, $f_sg$, while cells that have been starved will have less than this amount. If a cell is starved and then returned to excess fN, then its fN storage density will asymptotically approach $f_sg$.  

\begin{figure}[h]
 \begin{center}
    \includegraphics[width=3.0in]{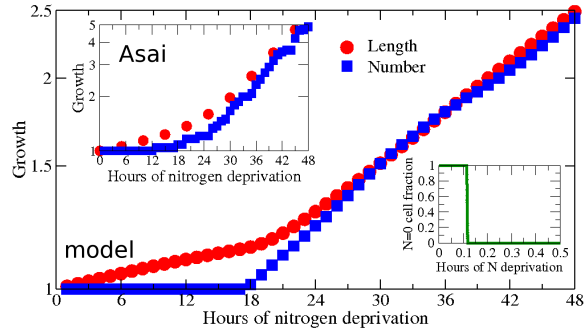}
  \end{center}
  \caption{\label{fig:growth} Growth of length (red circles) or number of cells (blue squares), relative to the initial values, of cyanobacterial filament vs.\ time since the start of fixed nitrogen deprivation. The top-left inset is experimental data digitized by hand from Movie~S1 in Asai \emph{et al} \cite{asai09}, while the main figure shows the computational model results. As described in the text, model division is suppressed when diffusible fixed nitrogen is unavailable --- though reduced growth is still allowed at the expense of fN storage.  The bottom-right inset shows the fraction of model cells with zero cytoplasmic, diffusible fixed nitrogen ($N_i=0$). Just after $t=0.1$hr the filament rapidly changes from all cells with $N_i>0$ to all cells with $N_i=0$. }
\end{figure}

In the model, the initial cytoplasmic fN concentration upon fN step-down is set so that cytoplasmic fN is exhausted in all cells within the first hour (see bottom-right inset of Fig.~\ref{fig:growth}). This is consistent with the activation of nitrate assimilation genes \cite{wolk96} and increase in 2-oxoglutarate levels \cite{laurent05} within 1h of nitrogen deprivation. Nevertheless, ongoing growth of cyanobacterial filaments is observed experimentally well after this time (see top-left inset of Fig.~\ref{fig:growth}) --- though it is reduced by a factor of $f_g \approx 0.25$ compared to later times after mature heterocysts are supplying newly fixed nitrogen \cite{asai09}. We find that utilization of fN storage is essential to recover this ongoing growth (see Fig.~\ref{fig:growth}). We also incorporate the experimental observation that division lags even behind the reduced growth in filament length for approximately the first 24h \cite{asai09} with a requirement that cells divide only when they have non-zero levels of cytoplasmic fN, i.e. $N_i>0$.  As shown in Fig.~\ref{fig:growth}, after mature heterocysts have developed and begin to produce fN at $\approx 18$h, division resumes and the filament resumes rapid growth.

%%%%%%%%%%%%%%%%%%%
\subsection{Estimation of storage fraction, $f_s$}
\label{sec:fs}
To estimate the storage fraction, $f_s$, we separately estimate the amounts of nitrogen contained in reserves of cyanophycin and phycobiliprotein in vegetative cells.  PCC 7120 has 636 $\mu$g arginine/mg chlorophyll, while a cyanobacterial species that does not synthesize cyanophycin has 68 $\mu$g arginine/mg chlorophyll, so we estimate that 568 $\mu$g arginine/mg chlorophyll is from cyanophycin \cite{picossi04}. Cyanophycin is 1:1 arginine:aspartic acid, where arginine has four nitrogen atoms and aspartic acid has only one, so there are approximately five nitrogen atoms in cyanophycin for every arginine from cyanophycin. \emph{Anabaena cylindrica} has 0.58 $\mu$g chlorophyll/10$^6$ cells, but its cells are 2.25 times as large as those of PCC 7120 \cite{brown12}, so we estimate that PCC 7120 has 0.26 $\mu$g chlorophyll/10$^6$ cells. Together,  568 arginine/mg chlorophyll $\times$ 5 N atoms per arginine $\times$ 0.26 $\mu$g chlorophyll/10$^6$ cells gives 2.55$\times$10$^9$ nitrogen atoms in cyanophycin per cell. There are 2.07$\times$10$^{10}$ nitrogen atoms per cell in PCC 7120  \cite{brown12} and so 12.3$\%$ of the nitrogen atoms are in cyanophycin. This ballpark estimate for the amount of cyanophycin is comparable to, though somewhat below, measurements of a maximum of 20$\%$ of protein in {\em Cyanothece} sp.\ ATCC 51142 \cite{li01}. Accordingly, we use a value of 15$\%$ of the fN in the cell in the form of cyanophycin.  Phycobiliprotein can comprise up to 60$\%$ of the soluble protein in a cell \cite{bogorad75} and upon nitrogen deprivation is observed to partially degrade in cells that remain vegetative \cite{bradley76}. We choose another 15$\%$ of the total fN in the cell to be storage in the form of phycobiliprotein that is available for growth upon nitrogen starvation. This gives a total of 30$\%$ of the fN as storage available for growth, and we assign $f_s=0.3$.   Other values of $f_s$ will lead to similar results, with the caveat that at least one of growth rate ($f_g \hat{R}$), fN requirements of growth ($g$), or the commitment threshold (see next section) would also need to be adjusted to retain the experimentally observed commitment timing. 

%%%%%%%%%%%%%%
\subsection{Heterocyst commitment}
The cyanobacterium \emph{Aphanocapsa}, when deprived of external fN, first consumes cyanophycin, then phycobiliprotein  \cite{allen84}. Vegetative cells only ever transiently and incompletely deplete their phycobiliprotein, as it participates in ongoing photosynthesis \cite{bradley76}. We follow this phenomenology and commit cells to become heterocysts when the available stored nitrogen is half of its maximal level, i.e. $N_{Si} < 0.5 f_s g l_i$ ($l$ is cell length) --- unless lateral inhibition (see below) prevents commitment.  Note that the precise threshold fraction used, like the fraction of fN represented by storage ($f_s$), does not change the qualitative results --- but does move the average timing of commitment earlier or later.  Nevertheless, commitment triggered at 50$\%$ of local storage roughly corresponds to local exhaustion of cyanophycin storage. Such local exhaustion may indicate a biological mechanism for heterocyst commitment.

\begin{figure}[t]
 \begin{center}
    \includegraphics[width=3.0in]{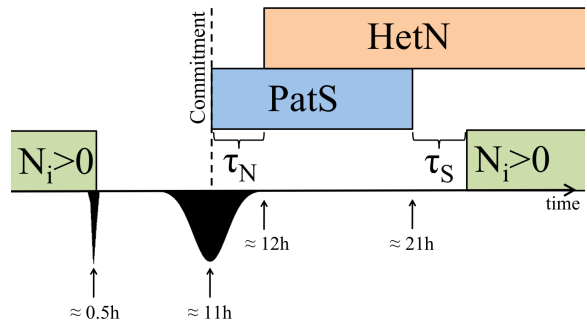}
  \end{center}
  \caption{\label{fig:5model} Heterocyst commitment and differentiation model for cells that commit to become heterocysts. Indicated times are measured from the onset of extracellular fixed nitrogen (fN) deprivation. Cytoplasmic fN (green bar on lower left) runs out relatively quickly, after less than 0.5h. In the model, the cell-to-cell variability of the timing of this depletion is relatively narrow, illustrated by the narrow black inverted peak at the lower left.  The cells then rely on stored fN for ongoing growth. Since stored fN is not shared between cells, it is significantly depleted at quite different times in different cells --- as represented by the broad black inverted peak centred at 11h.  When storage is significantly depleted, heterocyst commitment occurs --- as indicated by the vertical dashed line for a cell that commits at 11h.  Upon commitment, lateral inhibition due to {\em patS}  (blue bar) occurs immediately. Lateral inhibition due to {\em hetN} (orange bar) starts after a delay $\tau_N$ and is ongoing. PatS inhibition stops at heterocyst maturity (10h after commitment), and after a delay $\tau_S$ the mature heterocyst is fixing significant amounts of diffusible fN (green bar on lower right).  The inhibitory effects of PatS and HetN have a fixed range of five cells, in both directions, from the inhibiting cell.}
\end{figure}

Committed heterocysts do not grow in the model, i.e.\ $R_i=\hat{R}_i=0$. Rather they enter a deterministic developmental process. The stochastic nature of heterocyst commitment, in the model, arises entirely from the stochastic timing of stored nitrogen depletion (at approximately 11h, but broadly distributed, as illustrated by the broad inverted peak at the bottom of Fig.~\ref{fig:5model}).  We assume that after a cell commits to differentiation it will take 10 hours to mature developmentally (as indicated by the blue bar in Fig.~\ref{fig:5model}). This 10 hour interval is the delay seen in PCC 7120 between the 8-14 hour range of commitment \cite{yoon01} and the 18-24 hour expression of nitrogenase \cite{golden91}. This is also approximately the same delay seen in {\em Anabaena cylindrica} between the earliest commitment at 5h and the earliest mature heterocysts at 14h \cite{bradley76}. We also allow for a delay $\tau_S$ between maturity and significant production of fN . We explore the effects of varying $\tau_S$ (see results).  After this delay, mature heterocysts fix nitrogen at a rate $G_i=G_{het}=3.15\times10^6$ s$^{-1}$ \cite{brown12b}, as indicated by the green bar at the bottom right of Fig.~\ref{fig:5model}.

%%%%%%%%%%%%%%%%
\subsection{Lateral inhibition}
Heterocysts produce both PatS and HetN to laterally inhibit nearby cells from committing to become heterocysts. Both gene products are thought to be processed into a diffusible peptide \cite{yoon98, higa12}. {\em patS} acts early and affects the initial selection of cells for heterocyst development after fN deprivation \cite{yoon01, yoon98}. {\em hetN} acts later \cite{bauer97} and affects steady-state heterocyst patterns only after approximately one day of fN deprivation \cite{callahan01}. 

To highlight the role of fN storage, we have simplified the detailed dynamics of PatS and HetN \cite{zhu10} so that in the model the inhibition from a given cell is either on or off, i.e.\ it is represented by a Boolean model \cite{fisher07}. Similarly,  the inhibition effect is given a fixed range --- reflecting the fixed range found experimentally in filamentous cyanobacteria \cite{mitchison76}. For simplicity, and because inhibition by PatS and HetN may involve the same peptide \cite{risser09}, we use the same range for the effects of {\em patS} and {\em hetN}.  Reported experimental heterocyst frequencies at 24h include 7.8$\%$ \cite{sakr06}, 9.1$\%$ \cite{toyoshima10}, and 10.4$\%$ \cite{yoon01}. For the model, heterocyst frequency at 24h decreases as the inhibition range increases:  an inhibition range of 3 cells gives an 11.6$\%$ frequency, 4 cells 10.1$\%$, 5 cells 8.7$\%$, and 6 cells 7.6$\%$.  Accordingly, unless otherwise noted we choose an intermediate inhibition range of five, which also agrees with the range reported by Mitchison \emph{et al} \cite{mitchison76} for \emph{Anabaena catenula}. The effects of inhibition due to {\em patS} or {\em hetN} from a given cell are the same --- within the finite inhibition range no other cells will commit to becoming heterocysts. Those inhibited cells will continue to grow and divide, subject to available fN. Inhibition has no effect on already committed cells, or on mature heterocysts.

We have used a hierarchic \cite{buganim12}, or deterministic, model of inhibition timing.  As illustrated by the blue bar in Fig.~\ref{fig:5model}, the inhibition due to  {\em patS} begins at commitment and ends (10h later) at heterocyst maturity. Inhibition due to {\em hetN} is similar, but begins a time $\tau_N$ after commitment. Since the timing of commitment of individual cells can vary considerably, the timing of significant lateral inhibition from a given cell will correspondingly vary --- but the model has no stochasticity for the developmental process of an individual cell after commitment. 

There are three inhibitory products of committed cells. The first is PatS; after a delay $\tau_N$, the next is HetN; and a time $\tau_S$ after the {\em patS} turns off, newly synthesized fN inhibits commitment near the new heterocyst. In wild-type (WT) cells, these inhibitory signals overlap in time. Later, we will use the model to explore the effects of $\Delta patS$ and $\Delta hetN$ mutants in which inhibition of nearby cells is no longer uninterrupted after commitment.  We will also explore the hypothesis that heterocyst patterns similar to the recently reported $\Delta patN$ phenotype can be recovered by simply reducing the inhibition range.  All of the model results shown in this paper use short delays, with $\tau_N=\tau_S=1$h. We discuss (below) the sensitivity of the results on those delay times.

%%%%%%%%%%%%%%%%%
\subsection{Further numerical details}
For fixed nitrogen transport between cells, periodic boundary conditions were used to minimize end effects. Filaments were initiated as a single cell with zero cytoplasmic fN and with maximum storage, $N_S=f_s g l$, in a high concentration of external fN. After seven days of growth in high external fN, to generate a random population structure of cell lengths, the external fN concentration was stepped down (to zero unless otherwise stated) and the amount of cytoplasmic nitrogen was set to 5$\%$ of the current fN content incorporated into the cell, $0.05 g l_i$, so that the depletion of cytoplasmic fN is prior to the 0.5 hour time at which nitrate assimilation genes are activated \cite{wolk96} --- one of the earliest signs of nitrogen deprivation. 

Filament length and mature heterocyst frequency were recorded every hour. Heterocyst spacing distributions were recorded at the indicated times. Times are reported with respect to when the external fN concentration was stepped down. Commitment was measured similar to the experimental technique of Yoon and Golden \cite{yoon01}, where nitrate was added to the medium at a certain time and the subsequent heterocyst frequency at 24 hours is reported: in the simulation the external fN level was changed to a high concentration at the indicated time, and the filament allowed to grow until the heterocyst frequency was recorded at 24 hours.

%%%%%%%%%%%%%%%%%
%%%%%%%%%%%%%%%%%
\section{Results and Discussion}
\label{sec:results}

%%%%%%%%%%%%%%%%
\subsection{Commitment timing}

\begin{figure*}[t]
 \begin{center}
   \begin{tabular}{cc}
    \includegraphics[width=3.0in]{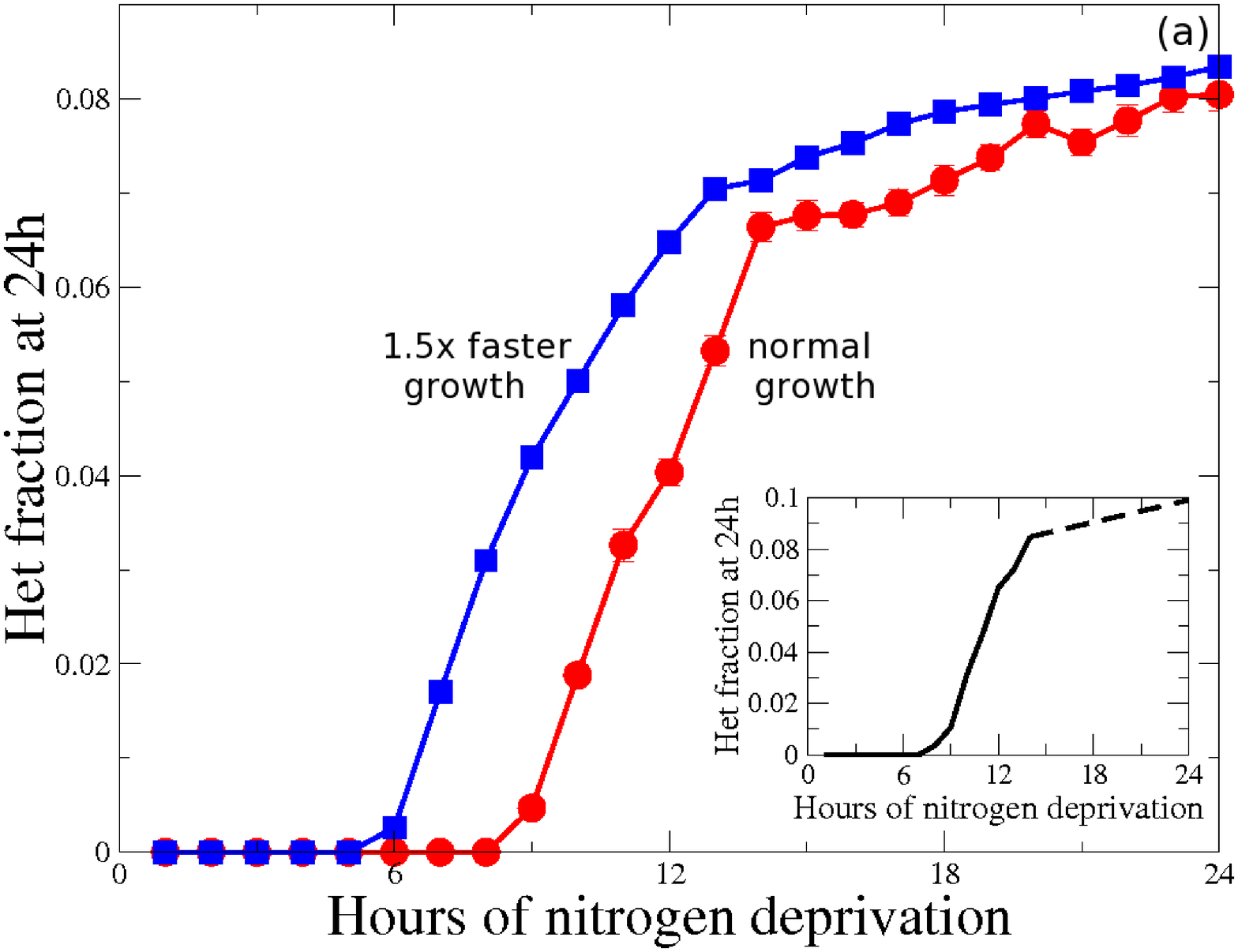}& \includegraphics[width=3.0in]{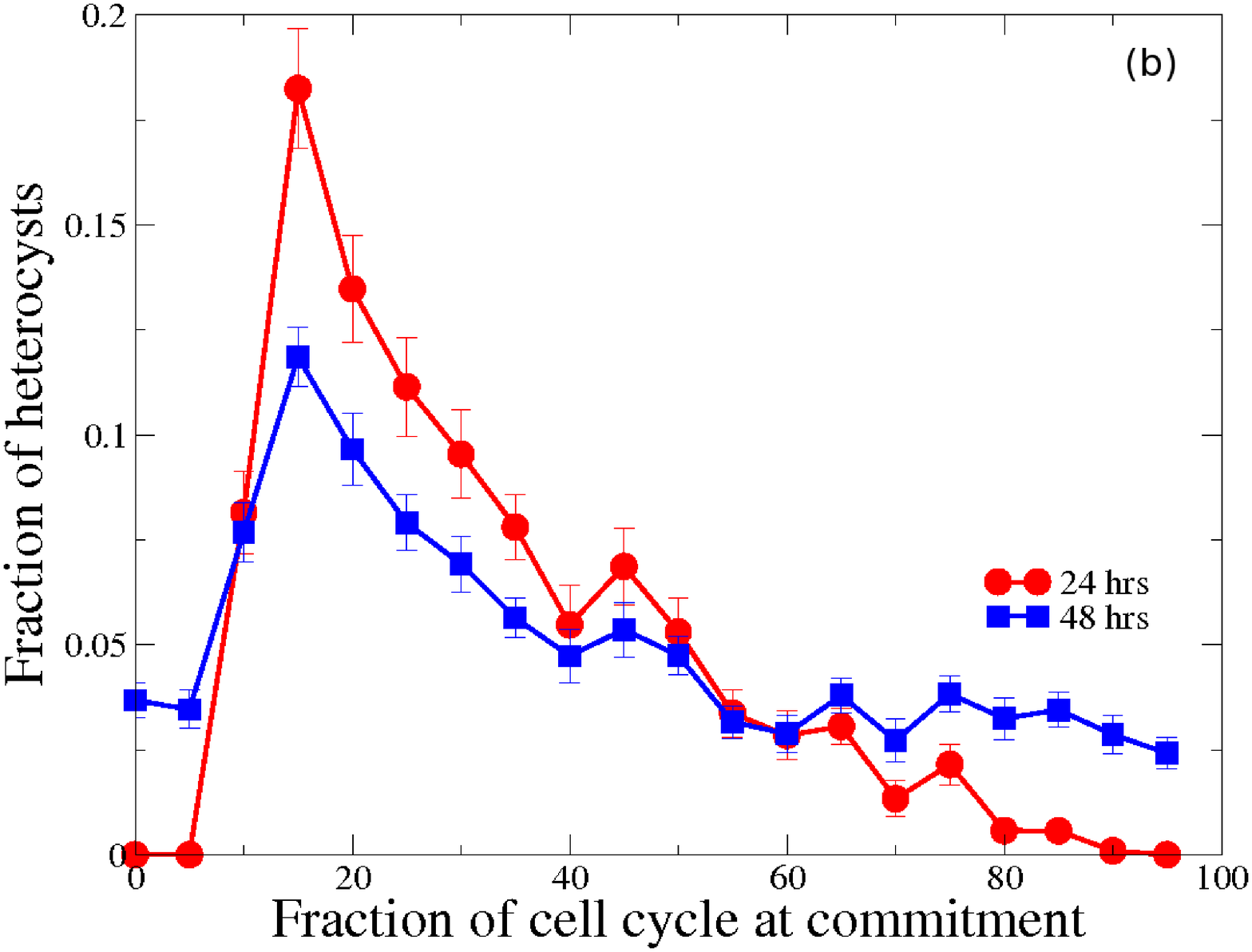}\\
    \includegraphics[width=3.0in]{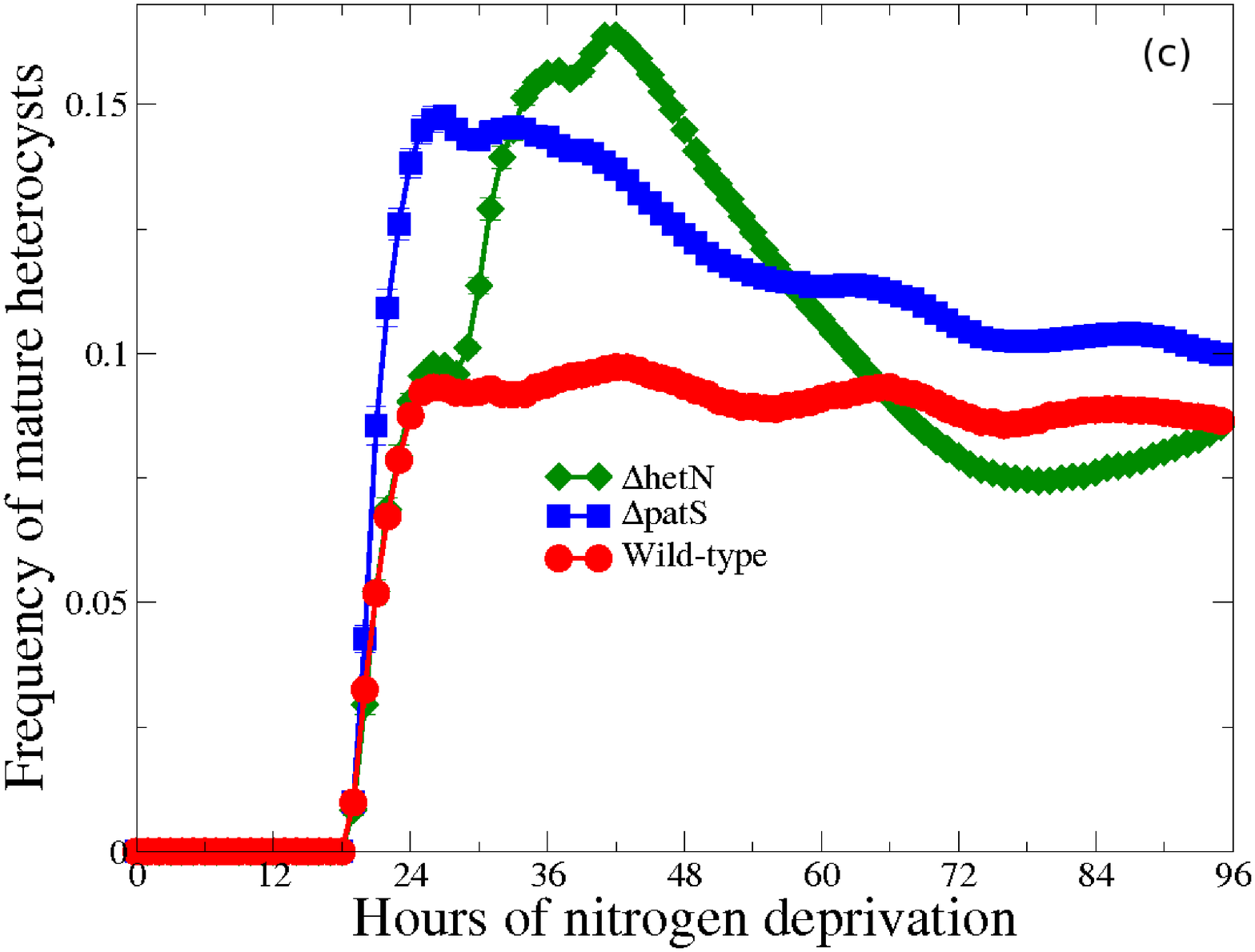}&  \includegraphics[width=3.0in]{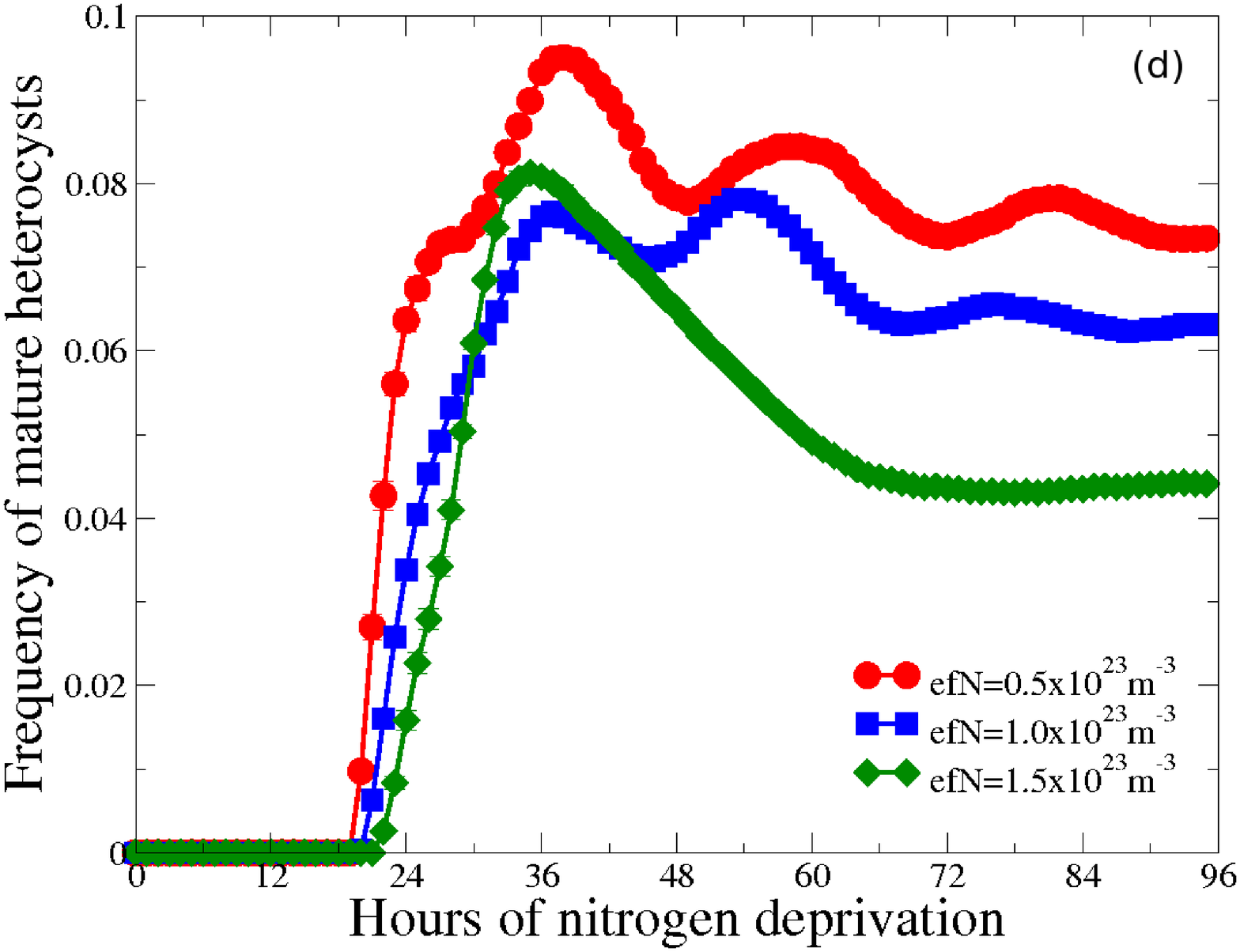} \\
  \end{tabular} 
  \end{center}
  \caption{\label{fig:checks}\label{fig:cellcycle} (a) Commitment curve of heterocyst fraction 24h after the onset of nitrogen deprivation vs.\  duration of nitrogen deprivation. Red circles are for a normal 20h doubling time, while the blue squares shows commitment fraction when the growth rate is increased by 50$\%$. Inset shows average of two sets of experimental data from Yoon and Golden \cite{yoon01} -- interpolated data to the point at 24h, representing uninterrupted deprivation, is shown with a dashed line. (b) Indirect cell-cycle effect, illustrated by the distribution of the cell sizes that heterocysts had at commitment. Legend indicates time since nitrogen deprivation, either 24h (red circles) or 48h (blue squares). (c) The change of heterocyst frequency with time after nitrogen deprivation for wild-type (red circles), $\Delta$\emph{patS} (blue squares), and $\Delta$\emph{hetN} (green diamonds) filaments. (d) The change of heterocyst frequency with time after nitrogen step-down for wild-type, with step-down to $\rho_{efN}$ (see Eqn.~\ref{eq:ndynamics}) of 0.5$\times$10$^{23}$m$^{-3}$ (red circles), 1.0$\times$10$^{23}$m$^{-3}$ (blue squares), and 1.5$\times$10$^{23}$m$^{-3}$ (green diamonds).}
\end{figure*}

The filament is deprived of external fN at zero hours. As the cells starve of fN some commit to develop into heterocysts --- following the deterministic developmental program outlined in Fig.~\ref{fig:5model}. Experimentally, commitment is assessed by raising the external fN to a high concentration after a delay \cite{yoon01,bradley76} and measuring the resulting heterocyst fraction at 24h. In PCC 7120, if the delay is shorter than around 8h no cells commit, while if longer than around 14h then the normal number of cells commit \cite{yoon01}.  As shown in Fig.~\ref{fig:checks}(a) with the red circles, we recover the same qualitative commitment curve with the model.

The variation of commitment timing is determined by the fN storage. In the model, commitment begins only after the first cells have depleted half of their stored fN. Because the maximum nitrogen storage is a fixed fraction $f_s$ of the cellular content, this occurs first in the smallest and in the fastest growing cells.  Cells will continue to commit until all cells have either committed to differentiation or been blocked by lateral inhibition. Because of the distribution of cellular stored fN and growth rates, commitment does not slow until about 14h. After that, there is little further commitment until the first mature heterocysts provide newly fixed nitrogen at 19h. 

With storage-based commitment we predict that filaments with faster growth rates will deplete their storage faster and commit to differentiation earlier. Bradley and Carr  found earlier commitment in \emph{Anabaena cylindrica}, ranging between 5-10h \cite{bradley76}.  Their doubling time for growth was 16.4h, shorter than a typical PCC 7120 generation time of about 20h \cite{picossi05}. As shown in Fig.~\ref{fig:checks}(a), we find that with a 50$\%$ faster growth rate the commitment curve started at 6h --- consistent with Bradley and Carr \cite{bradley76}.  It has also been observed experimentally that increasing light intensity, which results in increased growth rate, causes earlier heterocyst differentiation (see Fig.~S7 of \cite{toyoshima10}). Decreasing the amount of storage in our model produces a similar effect (see Fig.~S1 of this paper), which may also contribute to the earlier commitment seen in \emph{A. cylindrica}.

%%%%%%%%%%%%%%%
\subsection{Cell cycle effects}
While the cell cycle does not strictly regulate heterocyst commitment \cite{asai09, toyoshima10}, some association between the two has been reported \cite{mitchison76, sakr06}.  Since the model does not explicitly require that a cell is in a certain stage of the cell cycle to commit to heterocyst differentiation, we can use it to explore possible indirect effects of cell cycle on heterocyst commitment. 

Diffusible fN runs out for all cells in a narrow range of time (see Fig.~\ref{fig:growth}, bottom-right inset) because of the rapid rate of diffusion along the filament. After the diffusible fN has run out, all cells will grow using stored fN and smaller cells, having less storage, are on average expected to deplete storage sooner. Statistics on the length of the cell at the time of commitment are shown in Fig.~\ref{fig:cellcycle}(b). Cells double in length between birth and subsequent division: we take this fractional growth to define a cell-cycle between 0$\%$-100$\%$ extension. We see at 24h (red circles) that shorter cells, which are earlier in the cell cycle, are more likely to commit to heterocyst differentiation. This suggests there may be an indirect effect of cell cycle on heterocyst commitment. Interestingly, because even the shortest and fastest growing cells do not exhaust their storage immediately, the peak is not immediately after division. The cell-cycle effect is less pronounced at 48h, since only vegetative cells close to midway between existing heterocysts have a chance to exhaust their nitrogen storage \cite{brown12}.  

%%%%%%%%%%%%%%%%%%%%%%%
\subsection{Change of heterocyst frequency with time}
The frequency of mature heterocysts in wild-type filaments at different times after nitrogen deprivation, shown as the red circles of Fig.~\ref{fig:checks}(c), is similar to the commitment curve --- but delayed by 10h to allow for maturation. However, since exogenous fN is not reintroduced, the heterocyst frequency can be followed well after 24h. We see that the first heterocysts mature after 19h and the heterocyst frequency increases rapidly for several hours. This is consistent with the maturity of first generation heterocysts after 18-24h of nitrogen deprivation \cite{golden91, bohme88}. After 24h the heterocyst frequency oscillates, with a period of approximately 20 hours, while it slowly drops from a peak of nearly 10$\%$ at 42h to less than 9$\%$ at 96h. Both the oscillation and decline of the heterocyst frequency have been observed in  \emph{Anabaena cylindrica}  \cite{bradley76}.  In the model, the slight decline is due to dilution of the original burst of heterocysts with ongoing growth, while the oscillations are due to the periodic enrichment of small daughter cells produced in subsequent waves of cell division due to growth that follows the provision of fN by newly mature heterocysts.

%%%%%%%%%%%%%%%%%
\subsection{Heterocyst spacings}

\begin{figure}[t]
 \begin{center}
   	 \includegraphics[width=3.4in]{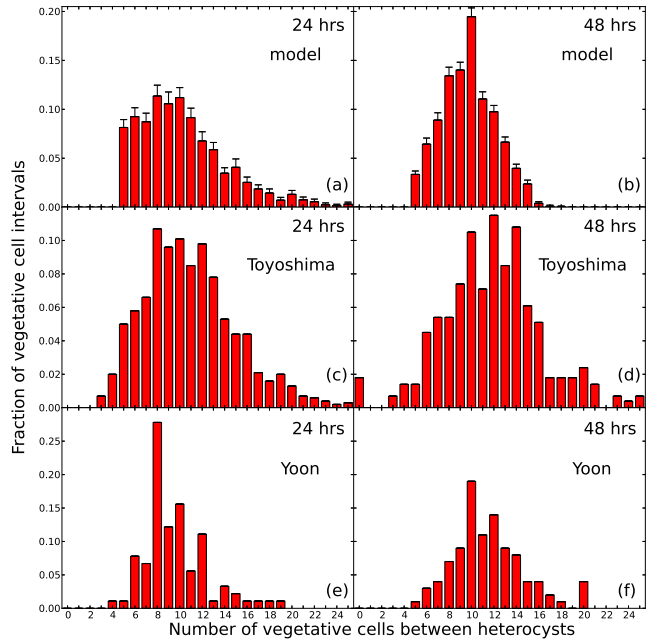}
  \end{center}
  \caption{\label{fig:spacings} Wild-type (WT) heterocyst spacing distributions. (a) and (b) spacing distribution at 24 hours and 48 hours, respectively, using the model.  Statistical error bars are shown. (c) and (d) spacing distribution from Toyoshima \emph{et al} \cite{toyoshima10} at 24 hours and 48 hours, respectively. (e) and (f) spacing distribution from Yoon and Golden \cite{yoon01} at 24 hours and 48 hours, respectively, where we note that spacings of 20 also include counts from all greater spacings.}
\end{figure}

The distribution of the number of cells separating mature heterocysts characterizes the developmental pattern. The heterocyst spacing distributions after 24 and 48h of nitrogen deprivation are shown for the model in Figs.~\ref{fig:spacings}(a) and (b) and for two experimental studies of PCC 7120 in Figs.~\ref{fig:spacings}(c)-(f).

The model distribution for 24h has a peak at 8-10 cells, and is skewed to the left. At higher spacings, the distribution tails off gradually. At smaller spacings, there is a sharp cutoff at the inhibition range. The sharpness of the cutoff is an artefact of the model's simple fixed inhibition range of five cells.  The spacing distribution evolves with time, and at 48h the mean spacing has increased and the distribution has become more symmetric. We see similar behavior in the experimental distributions: asymmetric left-skewed distributions at 24h, with smaller asymmetry but larger mean at 48h.  We also see considerable variation between the experimental distributions, indicating significant systematic variability of experimental conditions, which precludes a more detailed comparison with the model.

The model of commitment and lateral inhibition, Fig.~\ref{fig:5model}, captures both the distribution of spacing of heterocysts and the change of that distribution with time. The only source of stochasticity in the model is a modest variation of growth rates between cells \cite{brown12,allard07}, which is necessary to obtain a normal population structure of cells.  Other sources of stochasticity do not appear to be necessary to understand the broad, qualitative features of the heterocyst spacing distribution and its change with time.

%%%%%%%%%%%%%%%%%%%%%
\subsection{Heterocyst spacings with $\Delta$\emph{patS}}

\begin{figure}
 \begin{center}
   	 \includegraphics[width=3.4in]{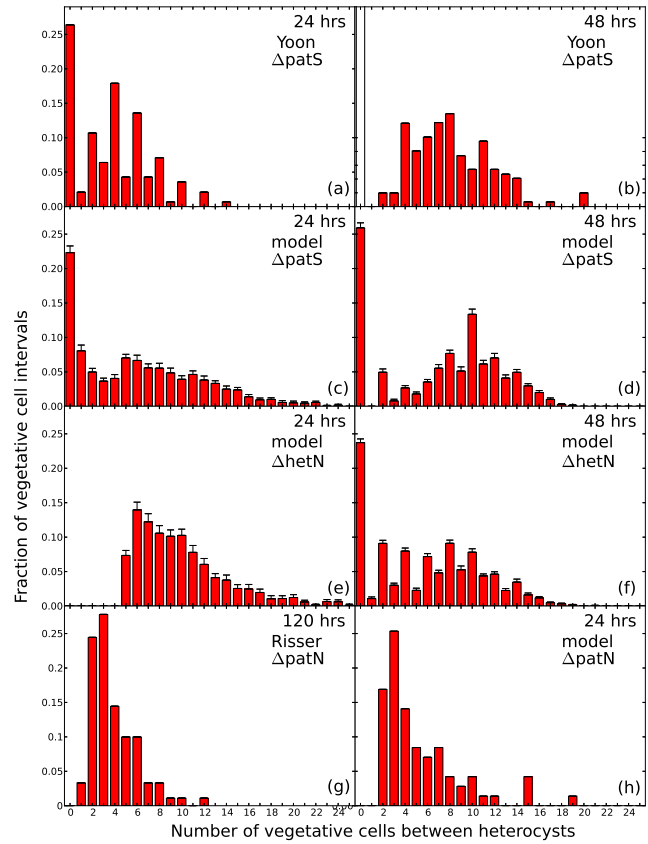}
  \end{center}
  \caption{\label{fig:pats} Heterocyst spacing distributions for inhibition mutants. Experimental $\Delta$\emph{patS} PCC 7120 from Yoon and Golden \cite{yoon01,yoon98} is shown in (a) and (b) with distributions at 24h and 48h as indicated. Experimental zero-spacing data at 48h was not reported \cite{yoon01}, so is represented by an unfilled bar. Model results for $\Delta$\emph{patS} are shown in (c) and (d). Model results for $\Delta$\emph{hetN} are shown in (e) and (f).  Experimental $\Delta$\emph{patN} \emph{Nostoc punctiforme} ATCC 29133 data from Fig.~S1(A) of Risser \emph{et al} \cite{risser12}  is shown in (g). Model results for an inhibition range of two cells are shown in (h).}
\end{figure}

Figs.~\ref{fig:pats}(a) and (b) show that experimental heterocyst spacing distributions without inhibition due to {\em patS} \cite{yoon01} are dramatically changed compared with wild-type (WT) experimental distributions. At 24h approximately 25$\%$ of the heterocysts spacings are zero, or Mch, and non-zero spacings are shorter overall \cite{yoon98}. At 48h the distribution of non-zero intervals is qualitatively similar to WT, however the fraction of Mch clusters was not reported and so is represented by an unfilled bar \cite{yoon01}.
  
Since local lateral inhibition is explicit in the model, we can investigate heterocyst patterns without {\em patS}.  We see from Fig.~\ref{fig:5model} that with $\Delta$\emph{patS}, committed heterocysts still laterally inhibit with {\em hetN} but only after a delay of duration $\tau_N$. We find (see Fig.~S2), that increasing $\tau_N$ increases the fraction of heterocyst clusters.  We use $\tau_N=1$h and recover the $\approx 25\%$ Mch fraction seen experimentally at 24h \cite{yoon01}.  The model results, in Figs.~\ref{fig:pats}(c) and (d), show the qualitative features seen experimentally: the shift towards lower spacings at 24h and the return of a qualitatively WT pattern of non-zero intervals at 48h.  The recovery of the distribution towards WT is partly caused by the production of fN by mature heterocysts. It is also due to HetN inhibition, which although later than the PatS inhibition, still restricts further commitment. Existing heterocyst clusters are diluted but not destroyed by ongoing growth and division of vegetative cells. The model indicates that more heterocysts commit after 24h, increasing the number of Mch clusters slightly more than they are diluted, so that at 48h the Mch fraction is still approximately 25$\%$.

The remaining heterocyst clusters also lead to an enhanced heterocyst frequency, as shown by blue squares in Fig.~\ref{fig:checks}(c).  The heterocyst frequency for $\Delta$\emph{patS} initially rises from zero at 19h, the same time as WT, but  peaks at a significantly higher frequency of 14.8$\%$ due to the lack of PatS inhibition. The percentage then slowly decreases towards the WT value, because the extra heterocysts limit nitrogen starvation and therefore new heterocyst differentiation.  Experimental heterocyst percentages vs.\ time for $\Delta$\emph{patS} \emph{Anabaena} PCC 7120 also show elevated heterocyst percentages ($\sim20\%$) that also appear to oscillate as time passes \cite{borthakur05}.

Experimental $\Delta$\emph{patS} heterocyst percentages are higher than those from the model. We believe that this is due to the absence of carbohydrate dynamics in the model,  so that when heterocyst percentages are elevated above WT we miss the expected reduction of fN production per heterocyst due to carbohydrate restriction \cite{murry89}.  This effect is also seen with $\Delta$\emph{hetN} strains (see next section).  We expect this to be an even more significant issue with the heterocyst proliferation seen in $\Delta$\emph{patS}$\Delta$\emph{hetN} double mutants \cite{borthakur05}, which we do not attempt to address without carbohydrate dynamics. 

%%%%%%%%%%%%%%%%%%%%%
\subsection{Heterocyst spacings with $\Delta$\emph{hetN}}
Experimental filaments without lateral inhibition due to \emph{hetN} exhibit  a wild-type heterocyst pattern at 24h, but a Mch phenotype at 48h \cite{callahan01}. 

As illustrated in Fig.~\ref{fig:5model}, local effects of $\Delta$\emph{hetN} in the model arise because of the gap of duration $\tau_S$ between the end of lateral inhibition due to {\em patS} and the start of lateral inhibition due to nitrogen fixation by the mature heterocyst.  Figs.~\ref{fig:pats}(e) and (f) show the model heterocyst spacing distributions at 24 and 48h for filaments without HetN type inhibition. At 24h the distribution is similar to the wild type distribution of Fig.~\ref{fig:spacings}(a), but at 48h we see a Mch phenotype as well as a significant number of shorter spacings between heterocysts. In Fig.~\ref{fig:checks}(c) the $\Delta$\emph{hetN} heterocyst frequency approximately follows the WT frequency until about 28h, when it begins to again rapidly increase to a peak of 16.4$\%$ at 42h. This large excess of heterocysts causes the frequency to subsequently drop to below the WT frequency before again increasing, showing an oscillation of much greater period and amplitude than either WT or $\Delta$\emph{patS}. Heterocyst frequency vs.\ time for \emph{Anabaena} PCC 7120 with \emph{hetN} transcription inactivated is very similar to wild-type for 24 hours, increasing to approximately 20$\%$ at 48 hours, and seems to slowly oscillate while increasing after 48 hours \cite{borthakur05}. The model has a smaller peak, and misses the subsequent slow increase, which we believe is due to its lack of carbohydrate dynamics. 

Variation of the duration of the inhibition gap $\tau_S$ did not significantly affect model results (see Fig.~S3, we use $\tau_S=\tau_N=1$h).  Essentially, any non-zero gap allows fN-depleted cells to commit. Despite the inhibition gap, newly committed cells will laterally inhibit their neighbours via {\em patS} --- so not all cells will commit, even though many vegetative cells have depleted nitrogen storage at 18-24h.  This leads to a significant difference between heterocyst clusters we see in $\Delta$\emph{patS} and those we see in $\Delta$\emph{hetN}. Heterocyst clusters in $\Delta$\emph{patS} occur because cells commit beside one another roughly at the same time --- neighbouring heterocysts are in the same generation. Conversely, heterocyst clusters in $\Delta$\emph{hetN} cannot arise this way because lateral inhibition due to {\em patS} prevents it. Rather, in $\Delta$\emph{hetN} heterocyst clusters arise because new heterocysts commit adjacent to older, mature heterocysts. 

%%%%%%%%%%%%%%%%%%%%%%%%%%%%%%
\subsection{Heterocyst spacings with $\Delta$\emph{patN}}  \label{sec:patn}
Filaments of \emph{Nostoc punctiforme} ATCC 29133 lacking \emph{patN} have an altered heterocyst pattern, with more heterocysts, as shown in Fig.~\ref{fig:pats}(g) \cite{risser12}.  While patterned expression of \emph{patN} may also influence the heterocyst pattern \cite{risser12}, we use the model to explore how much the reported heterocyst pattern may arise simply from decreasing the range of PatS and HetN inhibition.  

Fig.~S1(B) of Risser \emph{et al} \cite{risser12} indicates that significant nitrogen fixation and growth is only just starting at at 120h, when the heterocyst spacing distribution was measured.  In Fig.~\ref{fig:pats}(h) we show the heterocyst spacing distribution using the model for an inhibition range of two cells after 24h. This corresponds to when nitrogen fixation and growth starts again due to heterocyst maturation. We find that the heterocyst spacing distribution from the model with an inhibition range of two cells and the experimental $\Delta$\emph{patN} heterocyst spacing distribution are similar. This suggests that shortened inhibition ranges from PatS and HetN may play a role in shaping the $\Delta$\emph{patN} phenotype.

%%%%%%%%%
%%%%%%%%%
\subsection{Non-zero external fixed nitrogen concentrations}
Since local fN is an important part of local heterocyst placement in the model, we expect non-zero external fN levels to change both the heterocyst spacing distribution and heterocyst frequency. This was observed with a simpler computational model that was limited to steady-state growth, long after fN step-down \cite{brown12}.

Fig.~\ref{fig:cellcycle}(d), which shows the heterocyst frequency vs.\ time, illustrates that  strong relative differences can arise at around 24 hours due to slight differences in commitment timing. Fig.~\ref{fig:cellcycle}(d) also recovers the previous result \cite{brown12} that at later times the heterocyst frequency decreases as the external fN increases. At the highest external fN concentration shown, 1.5$\times10^{23}$m$^{-3}$, the late time heterocyst frequency is still nearly 5$\%$. In contrast, we find that at 2.0$\times10^{23}$m$^{-3}$, the heterocyst frequency does not rise above 1$\%$. 

\begin{figure}
 \begin{center}
   	 \includegraphics[width=3.4in]{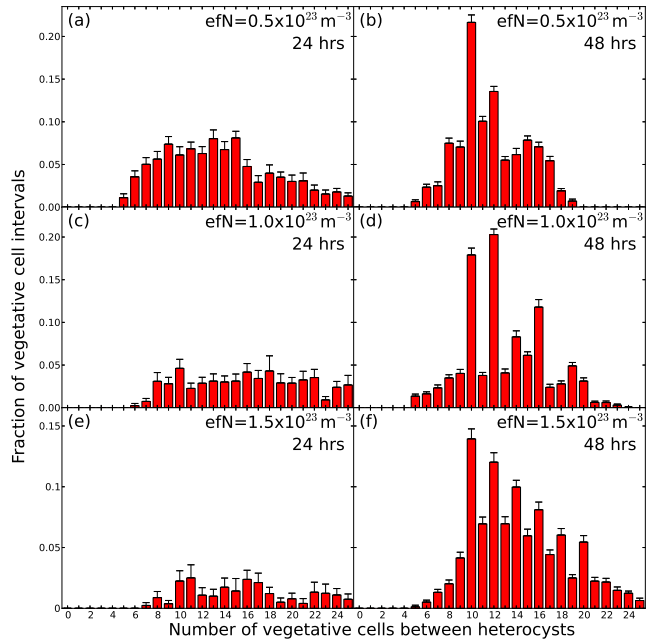}
  \end{center}
  \caption{\label{fig:extfnspacings} Wild-type heterocyst spacing distributions for different stepped down external fixed nitrogen (efN) concentrations. (a) and (b) spacing distribution at 24 hours and 48 hours, respectively, with efN concentration of $0.5\times$10$^{23}$m$^{-3}$ (see $\rho_{efN}$ in Eqn.~\ref{eq:ndynamics}). (c) and (d) spacing distribution at 24 hours and 48 hours, respectively, with efN concentration of $1.0\times$10$^{23}$m$^{-3}$. (e) and (f) spacing distribution at 24 hours and 48 hours, respectively, with efN concentration of $1.5\times$10$^{23}$m$^{-3}$.}
\end{figure}

We also observe that the oscillation of heterocyst numbers can lead to quite similar heterocyst percentages for significantly different levels of external fN. To illustrate this effect, in Fig.~\ref{fig:extfnspacings}, we show heterocyst spacing distributions from the model at 24 and 48 hours after fN step-down for three non-zero external fixed nitrogen concentrations. We see that relative differences in the spacing distributions are larger at 24h than at 48h, reflecting similar differences in heterocyst frequencies.

Thiel and Pratte \cite{thiel01} observed regular patterning of heterocysts in strains with vegetative cells expressing nitrogenase and producing fN under anoxic conditions. The nitrogenase in vegetative cells appeared after 2 hours of nitrogen deprivation, implying this fN was supplied prior to the typical commitment times that begin at 8 hours \cite{yoon01}. Since we find that  both heterocyst frequencies and heterocyst patterns do not dramatically change with moderate levels of non-heterocystous fN (in the case of the model, initially external to the filament), or even quite large levels of external fN at particular time points, we caution that the results of Thiel and Pratte may not rule out a storage-based heterocyst commitment model with a significant role for local fN in local heterocyst placement.

%%%%%%%%%
%%%%%%%%%
\section{Summary}

We have developed a dynamic developmental model of heterocystous cyanobacterial filaments that captures both the early and late heterocyst frequencies and spacings, in both wild-type filaments and in $\Delta${\em patS} and $\Delta${\em hetN} mutants. The computational model incorporates fixed nitrogen (fN) storage with fN transport, cell growth and division, and explicit lateral inhibition with heterocyst commitment and development.

Initial commitment of heterocysts is significantly delayed after the removal of fN in the extracellular medium \cite{yoon01, bradley76}, but the broad timing of commitment indicated by earlier experimental results shown in Fig.~\ref{fig:checks}(a) implies significant stochasticity in the commitment process. The model is deliberately as deterministic as possible; the only source of stochasticity is modest growth rate variability \cite{brown12,allard07} needed for a natural population structure that properly samples the entire range of cell sizes. Since the model recovers both heterocyst spacing and the distribution of the timing of commitment, we have shown that population context \cite{snijder09} is alone sufficient to explain observed patterns of heterocyst commitment. The commitment model is similar to a proposed two-stage model \cite{meeks02}, with  biased initiation relying upon storage levels rather than cell cycle position. 

Although it has been reported that only small newly divided daughter cells can differentiate into heterocysts \cite{mitchison76, sakr06, adams81},  recent reports are clear that division is not a requirement for heterocyst differentiation \cite{asai09,toyoshima10}.  Indeed, cell-lineage studies show that even inhibition of division leads to elongated heterocysts \cite{toyoshima10} rather than suppressing them altogether.  Nevertheless, it appears that cell-cycle and the cell-fate decision to develop into a heterocyst are correlated. 

Using our model, we examine the statistics of how heterocyst differentiation is related to cell size, since size is closely linked to the cell cycle in living cells. A strong indirect cell cycle commitment effect, seen in Fig.~\ref{fig:checks}(b), arises in the model from a uniform density of available storage determining a total storage that varies with cell size: smaller cells have less stored nitrogen. Because cell growth rates are not cell-size dependent in our model, smaller cells will typically exhaust their storage and commit earlier than larger cells. These cells then laterally inhibit other cells. This increased likelihood of small cells to differentiate into heterocysts due to more-rapid local depletion of storage offers an alternative explanation for a relationship between heterocyst differentiation and recent cell division. With the natural population structure of cell sizes in the filament, we believe that this explains the observed distribution of commitment times.  With lateral inhibition, this then allows different cell-fate decisions --- breaking the apparent initial symmetry of a clonal filament of vegetative cells.  This commitment model, in which commitment follows local storage depletion, predicts that the higher filamentous growth rate, as modulated by e.g. light intensity \cite{neunuebel08, toyoshima10}, should lead to earlier commitment, as  illustrated in Fig.~\ref{fig:checks}(a). We also expect that strains with less fN storage, such as strains without cyanophycin \cite{ziegler01}, with a correspondingly smaller $f_s$ in the model, should have significantly earlier commitment and correspondingly earlier heterocyst differentiation.

Lateral inhibition from committed and mature heterocysts, due to \emph{patS} \cite{yoon01,yoon98} and \emph{hetN} \cite{callahan01} respectively, is necessary to achieve the wild-type heterocyst pattern. Without {\em patS} inhibition that occurs upon commitment, cells can commit nearby one another and the 24h pattern exhibits Mch and very short spacings between heterocysts. Without {\em hetN} inhibition, the initial heterocyst pattern is qualitatively unchanged. However any temporal gap in the inhibition from a maturing heterocyst between {\em patS} and newly synthesized fN ($\tau_S$ in Fig.~\ref{fig:5model}) will lead to Mch and short spacings at 48h. Because {\em patS} continues to prevent simultaneous commitment of new heterocyst clusters, the Mch clusters seen in $\Delta${\em hetN} are always built around older heterocysts in the model. This is both a novel prediction of, and a natural consequence of, the model. We also found that reducing the inhibition range of \emph{patS} and \emph{hetN} to two cells yielded a heterocyst pattern similar to that of $\Delta$\emph{patN} \cite{risser12} filaments (see Sec.~\ref{sec:patn} and Figs.~\ref{fig:pats}(g) and (h)).

Motivated by the observation of qualitatively unchanged heterocyst development despite ongoing nitrogen fixation in vegetative cells under anoxic conditions \cite{thiel01}, we investigated step-down to nonzero external fN concentrations. After stepping down to nonzero external fN concentrations, no dramatic changes were seen in the modelled heterocyst pattern, frequency, and differentiation timing until the external fN nearly supressed heterocysts altogether. This is particularly true at intermediate times after fN step-down. This suggests that our model of heterocyst commitment due to local fN levels may be consistent with the experimental observation of heterocysts despite nitrogen fixation in vegetative cells under anoxic conditions.  Quantitative agreement of model and experimental patterns will require characterization of fN sources, sinks, transport and storage, as well as stochastic cellular growth.

As more detailed spatial and temporal patterns of expression of patterning genes are forthcoming, they can be more accurately represented in the model. Some refinements of the model are already suggested from the artificially sharp inhibition range built into the model and evident in Fig.~\ref{fig:spacings} compared with experimental short-range spacings. We have also noted that carbohydrate dynamics, with sources in vegetative cells and sinks in heterocysts, may be needed when heterocyst frequencies exceed those of WT.

The developmental model, summarized in  Fig.~\ref{fig:5model}, is deliberately simplified. It is deterministic, or hierarchical \cite{buganim12}, after initial commitment. Subsequent lateral inhibition is Boolean (on or off) in character \cite{fisher07}, and of fixed range. We know that the details of the biological regulatory network of lateral inhibition and commitment are much more complex \cite{flores10, kumar10}. Nevertheless, while many ``patterning'' genes have been identified that change heterocyst patterns when perturbed, their mechanism of action is sometimes less clear \cite{flores10}. We believe that the less-visible landscape of metabolites must be taken into account while determining mechanisms of patterning gene action.  For the model of heterocyst development in cyanobacterial filaments, we emphasize the role of fN and especially of fN storage in determining the observed developmental pattern.  This is consistent with the increasingly prominent role played by metabolic gradients in determining patterns of multicellular development in  bacterial systems (see, e.g., \cite{serra13}).

Nevertheless, a key question is how the expression of patterning genes correlates with subsequent heterocyst development, both spatially and temporally.  In particular, how variable is the timing and strength of expression of inhibition factors from cells that subsequently develop into heterocysts? The model, with hierarchic timing and Boolean expression after commitment, can explain observed heterocyst patterns.  Whether this indicates similarly hierarchic timing {\em in vivo}, or simply robustness of the spacing distributions to additional experimental stochasticities, remains to be determined.

There remains considerable experimental uncertainty in almost all of the parameters used in the model. We have tried to explicitly avoid fine-tuning of parameters by being clear about the reasoning and experimental data involved in determining the parameters. Nevertheless, appropriate model parameters are expected to depend on experimental growth conditions, as well as the cyanobacterial species and strain or mutant used. A detailed understanding of phenotypic heterocyst patterning differences with varying experimental conditions will probably require identification and measurement of the most significant parameter differences. We expect that growth rates and nitrogen storage dynamics will be among these significant parameters, and that such parameters should be characterized whenever possible. 

\ack
We thank the Natural Science and Engineering Research Council (NSERC) for support, and the Atlantic Computational Excellence Network (ACEnet) for computational resources. AIB also thanks NSERC, ACEnet, the Sumner Foundation, and the Killam Trusts for fellowship support.

%%%%%%%%%%
%%%%%%%%%%
\pagebreak

%recommended to use unsrt for some reason 
\section*{References}
\bibliographystyle{unsrt}
\bibliography{references}

\begin{thebibliography}{10}

\bibitem{flores10}
Flores E and Herrero A.
\newblock Compartmentalized function through cell differentiation in
  filamentous cyanobacteria.
\newblock {\em Nat Rev Microbiol}, 8:39--50, 2010.

\bibitem{kumar10}
Kumar K, Mella-Herrera~R A, and Golden~J W.
\newblock Cyanobacterial heterocysts.
\newblock {\em Cold Spring Harb Perspect Biol}, 2:a000315, 2010.

\bibitem{wolk96}
Wolk~C P.
\newblock Heterocyst formation.
\newblock {\em Annu Rev Genet}, 30:59--78, 1996.

\bibitem{laurent05}
Laurent S, Chen H, Bedu S, Ziarelli F, Peng L, and Zhang C.
\newblock Nonmetabolizable analogue of 2-oxoglutarate elicits heterocyst
  differentiation under repressive conditions in \emph{{A}nabaena} sp.\ {PCC}
  7120.
\newblock {\em Arch Microbiol}, 176:9--18, 2001.

\bibitem{yoon01}
Yoon H and Golden~J W.
\newblock Pat{S} and products of nitrogen fixation control heterocyst pattern.
\newblock {\em J Bacteriol}, 183:2605--2613, 2001.

\bibitem{mullineaux08}
Mullineaux~C W, Mariscal V, Nenninger A, Khanum H, Herrero A, Flores E, and
  Adams~D G.
\newblock Mechanism of intercellular molecular exchange in heterocyst-forming
  cyanobacteria.
\newblock {\em EMBO J}, 27:1299--1308, 2008.

\bibitem{brown12}
Brown~A I and Rutenberg~A D.
\newblock Reconciling cyanobacterial fixed-nitrogen distributions and transport
  experiments with quantitative modelling.
\newblock {\em Phys Biol}, 9:016007, 2012.

\bibitem{mitchison76}
Mitchison~G J, Wilcox M, and Smith~R J.
\newblock Measurement of an inhibitory zone.
\newblock {\em Science}, 191:866--868, 1976.

\bibitem{sakr06}
Sakr S, Jeanjean R, Zhang C, and Arcondeguy T.
\newblock Inhibition of cell division suppresses heterocyst development in
  \emph{{A}nabaena} sp. strain {PCC} 7120.
\newblock {\em J Bacteriol}, 188:1396--1404, 2006.

\bibitem{asai09}
Asai H, Iwamori S, Kawai K, Ehira S, Ishihara J, Aihara K, Shoji S, and Iwasaki
  H.
\newblock Cyanobacterial cell lineage analysis of the spatiotemporal
  \emph{het{R}} expression profile during heterocyst pattern formation in
  \emph{{A}nabaena} sp. {PCC} 7120.
\newblock {\em PLOS One}, 4:e7371, 2009.

\bibitem{toyoshima10}
Toyoshima M, Sasaki~N V, Fujiwara M, Ehira S, Ohmori M, and Sato N.
\newblock Early candidacy for differentiation into heterocysts in the
  filamentous cyanobacterium \emph{{A}nabaena} sp. {PCC} 7120.
\newblock {\em Arch Microbiol}, 192:23--31, 2010.

\bibitem{nachman07}
Nachman I, Regev A, and Ramanathan S.
\newblock Dissecting timing variability in yeast meiosis.
\newblock {\em Cell}, 131:544--556, 2007.

\bibitem{stpierre08}
St-Pierre F and Endy D.
\newblock Determination of cell fate selection during phage lambda infection.
\newblock {\em Proc Natl Acad Sci USA}, 105:20705--20710, 2008.

\bibitem{li01}
Li~H, Sherman~D M, Bao S, and Sherman~L A.
\newblock Pattern of cyanophycin accumulation in nitrogen-fixing and
  non-nitrogen-fixing cyanobacteria.
\newblock {\em Arch Microbiol}, 176:9--18, 2001.

\bibitem{bogorad75}
Bogorad L.
\newblock Phycobiliproteins and complementary chromatic adaptation.
\newblock {\em Ann Rev Plant Physiol}, 26:369--401, 1975.

\bibitem{allen84}
Allen~M M.
\newblock Cyanobacterial cell inclusions.
\newblock {\em Ann Rev Microbiol}, 38:1--25, 1984.

\bibitem{buikema93}
Buikema~W J and Haselkorn R.
\newblock Molecular genetics of cyanobacterial development.
\newblock {\em Ann Rev Plant Physiol Plant Mol Biol}, 44:33--52, 1993.

\bibitem{buganim12}
Buganim Y, Faddah~D A, Cheng~A W, Itskovich E, Markoulaki S, Ganz K, Klemm~S L,
  van Oudenaarden~A, and Jaenisch R.
\newblock Single-cell expression analyses during cellular reprogramming reveal
  an early stochastic and a late hierarchic phase.
\newblock {\em Cell}, 150:1209--1222, 2012.

\bibitem{yoon98}
Yoon H and Golden~J W.
\newblock Heterocyst pattern formation controlled by a diffusible peptide.
\newblock {\em Science}, 282:935--938, 1998.

\bibitem{callahan01}
Callahan~S M and Buikema~W J.
\newblock The role of {H}et{N} in maintenance of the heterocyst pattern in
  \emph{{A}nabaena} sp. {PCC} 7120.
\newblock {\em Mol Microbiol}, 40:941--950, 2001.

\bibitem{snijder09}
Snijder B, Sacher R, Ramo P, Damm E, Liberali P, and Pelkmans L.
\newblock Population context determines cell-to-cell variability in endocytosis
  and virus infection.
\newblock {\em Nature}, 461:520--523, 2009.

\bibitem{risser12}
Risser~D D, Wong F~C Y, and Meeks~J C.
\newblock Biased inheritance of the protein {P}at{N} frees vegetative cells to
  initiate patterned heterocyst differentiation.
\newblock {\em Proc Natl Acad Sci USA}, 109:15342--15347, 2012.

\bibitem{thiel01}
Thiel T and Pratte B.
\newblock Effect on heterocyst differentiation of nitrogen fixation in
  vegetative cells of the cyanobacterium \emph{{A}nabaena variabilis} {ATCC}
  29413.
\newblock {\em J Bacteriol}, 183:280--286, 2001.

\bibitem{wolk74}
Wolk~C P, Austin~S M, Bortins J, and Galonsky A.
\newblock Autoradiographic localization of $^{13}${N} after fixation of
  $^{13}${N}-labeled nitrogen gas by a heterocyst-forming blue-green alga.
\newblock {\em J Cell Biol}, 61:440--453, 1974.

\bibitem{allard07}
Allard~J F, Hill~A L, and Rutenberg~A D.
\newblock Heterocyst patterns without patterning proteins in cyanobacterial
  filaments.
\newblock {\em Dev Biol}, 312:427--434, 2007.

\bibitem{brown12b}
Brown~A I and Rutenberg~A D.
\newblock Heterocyst placement strategies to maximize the growth of
  cyanobacterial filaments.
\newblock {\em Phys Biol}, 9:046002, 2012.

\bibitem{meeks02}
Meeks~J C and Elhai J.
\newblock Regulation of cellular differentiation in filamentous cyanobacteria
  in free-living and plant-associated symbiotic growth states.
\newblock {\em Microbiol Mol Biol Rev}, 66:94--121, 2002.

\bibitem{wolk75}
Wolk~C P and Quine~M P.
\newblock Formation of one-dimensional patterns by stochastic processes and by
  filamentous blue-green algae.
\newblock {\em Dev Biol}, 46:370--382, 1975.

\bibitem{zhu10}
Zhu M, Callahan~S M, and Allen~J S.
\newblock Maintenance of heterocyst patterning in a filamentous cyanobacterium.
\newblock {\em J Biol Dyn}, 4:621--633, 2010.

\bibitem{picossi05}
Picossi S, Montesinos~M L, Pernil R, Lichtle C, Herrero A, and Flores E.
\newblock {ABC}-type neutral amino acid permease {N}-{I} is required for
  optimal diazotrophic growth and is repressed in the heterocysts of
  \emph{{A}nabaena} sp. strain {PCC} 7120.
\newblock {\em Mol Microbiol}, 57:1582--1592, 2005.

\bibitem{reshes08}
Reshes G, Vanounou S, Fishov I, and Feingold M.
\newblock Cell shape dynamics in \emph{{E}scherichia coli}.
\newblock {\em Biophys J}, 94:251--264, 2008.

\bibitem{picossi04}
Picossi S, Valladares A, Flores E, and Herrero A.
\newblock Nitrogen-regulated genes for the metabolism of cyanophycin, a
  bacterial nitrogen reserve polymer.
\newblock {\em J Biol Chem}, 279:11582--11592, 2004.

\bibitem{bradley76}
Bradley S and Carr~N G.
\newblock Heterocyst and nitrogenase development in \emph{{A}nabaena
  cylindrica}.
\newblock {\em J Gen Microbiol}, 96:175--184, 1976.

\bibitem{golden91}
Golden~J W, Whorff~L L, and Wiest~D R.
\newblock Independent regulation of \emph{nif{HDK}} operon transcription and
  {DNA} rearrangement during heterocyst differentiation in the cyanobacterium
  \emph{{A}nabaena} sp. strain {PCC} 7120.
\newblock {\em J Bacteriol}, 173:7098--7105, 1991.

\bibitem{higa12}
Higa~K C, Rajagopalan R, Risser~D D, Rivers~O S, Tom~S K, Videau P, and
  Callahan~S M.
\newblock The {RGSGR} amino acid motif of the intercellular signalling protein,
  {H}et{N}, is required for patterning of heterocysts in \emph{{A}nabaena} sp.
  strain {PCC} 7120.
\newblock {\em Mol Microbiol}, 83:682--693, 2012.

\bibitem{bauer97}
Bauer~C C, Ramaswamy~K S, Endley S, Scappino~L A, Golden~J W, and Haselkorn R.
\newblock Suppression of heterocyst differentiation in \emph{{A}nabaena} {PCC}
  7120 by a cosmid carrying wild-type genes encoding enzymes for fatty acid
  synthesis.
\newblock {\em FEMS Microbiol Lett}, 151:23--30, 1997.

\bibitem{fisher07}
Fisher J and Henzinger~T A.
\newblock Executable cell biology.
\newblock {\em Nat Biotechnol}, 25:1239--1249, 2007.

\bibitem{risser09}
Risser~D D and Callahan~S M.
\newblock Genetic and cytological evidence that heterocyst patterning is
  regulated by inhibitor gradients that promote activator decay.
\newblock {\em Proc Natl Acad Sci USA}, 106:19884--19888, 2009.

\bibitem{bohme88}
Bohme H and Haselkorn R.
\newblock Molecular cloning and nucleotide sequence analysis of the gene coding
  for heterocyst ferrodoxin from the cyanobacterium \emph{{A}nabaena} sp.
  strain {PCC} 7120.
\newblock {\em Mol Gen Genet}, 214:278--285, 1988.

\bibitem{borthakur05}
Borthakur~P B, Orozco~C C, Young-Robbins~S S, Haselkorn R, and Callahan~S M.
\newblock Inactivation of \emph{pat{S}} and \emph{het{N}} causes lethal levels
  of heterocyst differentiation in the filamentous cyanobacterium
  \emph{{A}nabaena} sp. {PCC} 7120.
\newblock {\em Mol Microbiol}, 57:111--123, 2005.

\bibitem{murry89}
Murry~M A and Wolk~C P.
\newblock Evidence that the barrier to the penetration of oxygen into
  heterocysts depends upon two layers of the cell envelope.
\newblock {\em Arch Microbiol}, 151:469--474, 1989.

\bibitem{adams81}
Adams~D G and Carr~N G.
\newblock The developmental biology of heterocyst and akinete formation in
  cyanobacteria.
\newblock {\em Crit Rev Microbiol}, 9:45--100, 1981.

\bibitem{neunuebel08}
Neunuebel~M R and Golden~J W.
\newblock The \emph{{A}nabaena} sp. strain {PCC} 7120 gene all2874 encodes a
  diguanylate cyclase and is required for normal heterocyst development under
  high-light growth conditions.
\newblock {\em J Bacteriol}, 190:6829--6836, 2008.

\bibitem{ziegler01}
Ziegler K, Stephan~D P, Pistorius~E K, Ruppel~H G, and Lockau W.
\newblock A mutant of the cyanobacterium \emph{{A}nabaena variabilis} {ATCC}
  29413 lacking cyanophycin synthetase: growth properties and ultrastructural
  aspects.
\newblock {\em FEMS Microbiol Lett}, 196:13--18, 2001.

\bibitem{serra13}
Serra~D O, Richter~A M, Klauck G, Mika F, and Hengge R.
\newblock Microanatomy at cellular resolution and spatial order of
  physiological differentiation in a bacterial biofilm.
\newblock {\em mBio}, 4:e00103--13, 2013.

\end{thebibliography}

\end{document}